# Entropy on the von Neumann lattice and its evaluation

Sumiyoshi Abe[†] and J. Zak[‡]

[†]Institute of Physics, University of Tsukuba, Ibaraki 305-8571, Japan

[‡]Department of Physics, Technion-Israel Institute of Technology,

 Haifa 32000, Israel

**Abstract**. Based on the recently introduced averaging procedure in phase space, a new type of entropy is defined on the von Neumann lattice. This quantity can be interpreted as a measure of uncertainty associated with simultaneous measurement of the position and momentum observables in the discrete subset of the phase space. Evaluating for a class of the coherent states, it is shown that this entropy takes a stationary value for the ground states modulo of a unit cell of the lattice in such a class. This value for the ground state depends on the ratio of the position lattice spacing and the momentum lattice spacing. It is found that its minimum is realized for the perfect square lattice, i.e., absence of squeezing. Numerical evaluation of this minimum gives  1.386....

Ever since the introduction of the Wigner distribution function [1], theory of phase-space representations [2-4] has been playing vital roles in quantum statistical mechanics, quantum chaos and signal processing. The Husimi function, which is the coherent-state representation of a wave function (or a density matrix) and is often called the *Q*-function, is known to be special in the sense that it is nonnegative and has a good analytical behavior in contrast to the Wigner function and the Sudarshan-Glauber *P*-function. Since the coherent state of the harmonic oscillator is the most classical pure state with the minimum Heisenberg uncertainty product, the Husimi *Q*-function is supposed to describe a semiclassical aspect of the system under consideration. However, a measure of classical nature of a normalized state, $|\psi\rangle$, was lacking until Wehrl [5] introduced a classical entropy, which is defined by

$$W[\psi] = -\iint d^2\alpha \frac{|\langle\alpha|\psi\rangle|^2}{\pi} \ln \frac{|\langle\alpha|\psi\rangle|^2}{\pi}. \tag{1}$$

$|\alpha\rangle$ in this equation is the coherent state given by $|\alpha\rangle = \hat{D}(\alpha)|0\rangle$ $(\alpha \in \mathbb{C})$, where

$$\hat{D}(\alpha) = \exp\left(\alpha \hat{a}^\dagger - \alpha^* \hat{a}\right) \tag{2}$$

is the displacement operator, $|0\rangle$ the normalized ground state of the harmonic oscillator, and $\hat{a}^\dagger$ and $a$ respectively the creation and annihilation operators given in terms of the position and momentum operators, $\hat{x}$ and $\hat{p}$, by

$$\hat{a}^{\dagger} = \frac{1}{\lambda\sqrt{2}}\left(\hat{x} - i\frac{\lambda^2}{\hbar}\hat{p}\right), \quad \hat{a} = \frac{1}{\lambda\sqrt{2}}\left(\hat{x} + i\frac{\lambda^2}{\hbar}\hat{p}\right), \tag{3}$$

where $\lambda$ is a constant and has often the meaning of the width of the harmonic oscillator. A pure state is considered here, but it is straightforward to generalize Eq. (1) to the case of a mixed state by simply making replacement $|\langle\alpha|\psi\rangle|^2 \to \langle\alpha|\hat{\rho}|\alpha\rangle$, where $\hat{\rho}$ is the density matrix. Note that Eq. (1) is different from Wehrl's original definition by the factor $1/\pi$ inside the logarithm. We introduce this factor here since the normalization condition on $|\psi\rangle$ is $\iint d^2\alpha \, |\langle\alpha|\psi\rangle|^2 / \pi = 1$.

Wehrl [5] has evaluated this entropy for a class of the coherent states $\{|z\rangle = \hat{D}(z)|0\rangle \, | \, z \in \mathbb{C}\}$ and has found that $W[z] = 1 + \ln\pi$. Then, he has conjectured the inequality:

$$W[\psi] \geq 1 + \ln\pi. \tag{4}$$

This conjecture has immediately been proved in Ref. [6] affirmatively.

Here, we note that the fact that $W[z]$ is independent of the coherent-state parameter, $z$, is due to the translational invariance of the Wehrl entropy, and therefore $W[z]$ is essentially the value for the ground state, $|0\rangle$.

The Wehrl entropy has repeatedly been used in the literature. For example, in Ref. [7] this entropy and the corresponding marginal entropies are considered in order to study the statistical properties of the squeezed states. In Ref. [8] it is employed as a sampling entropy associated with operational measurements of quantum states, in which the ruler states are chosen to be the coherent states. Another example is found in Ref. [9], where the problem of discriminating quantum states in decoherence processes is discussed by using this quantity.

The Wehrl entropy is defined in terms of the coherent-state representation of a quantum state. In Ref. [10] it has been discussed in detail that the coherent-state representation may be interpreted as simultaneous measurement of the position and momentum observables, or a pair of the quadrature operators in quantum optics. Therefore, the Wehrl entropy can also be regarded as a measure of uncertainty associated with such measurement.

As is well known, the coherent states possess the over-completeness property. However, as noticed by von Neumann [11], a discrete subset of the coherent states, in which a single state is assigned to a unit cell of area $2\pi\hbar$ in the phase plane, still forms a complete system [12-14]. (Strictly speaking, the von Neumann lattice states are still over-complete by one state.) Nowadays this subset is called the von Neumann lattice.

Therefore, to maintain informational completeness of a given quantum state, it is sufficient to consider it only on the von Neumann lattice. Actually, it is of physical relevance to process the state on the lattice since in practice one often uses a (finite) set of detectors arranged on the lattice points to perform measurements [15]. Thus, considering the state on the von Neumann lattice is equivalent to performing discrete simultaneous measurements of the position and momentum observables.

Until very recently, this program could not be carried out, however. This is because it was not known how to construct the probability distribution function associated with the state $|\psi\rangle$ on the von Neumann lattice. A main difficulty comes from nonorthogonality of the coherent states and thus the von Neumann lattice states. Of course, one could imagine that the standard orthogonalization procedure would be applied to the coherent states. However, then the Balian-Low theorem [16,17] tells us that in such orthogonalized states the Heisenberg uncertainty product $\Delta X \cdot \Delta P$ is divergent, destroying classical feature of the coherent states. Recently, this point has been overcome in Ref. [18]. Based on the fact that there are an infinite number of the coherent states in a unit cell of the lattice, the procedure of averaging the coherent states over the unit cell has been developed to construct an orthonormal system on the lattice.

In this paper, we present a new type of entropy, which is defined on the von Neumann lattice through averaging the Husimi $Q$-function over a unit cell. This entropy can be thought of as a measure of uncertainty associated with discrete simultaneous measurements of the position and momentum observables on the lattice points. We evaluate it for a class of the coherent states and show that it takes a stationary value for the ground states modulo of a unit cell of the von Neumann lattice in this class. Due to the lattice structure, the continuous translation invariance is broken, in contrast to the Wehrl entropy. An interesting point is that this value for the ground states explicitly depends on the ratio of the von Neumann lattice constant to $\lambda$ in Eq. (3). We find that this stationary value becomes the minimum for the perfect square lattice, i.e., absence of squeezing. Therefore, in this sense, this entropy can also be regarded as a measure of classical nature of a quantum state, like the Wehrl entropy.

Let us first summarize the averaging procedure developed in Ref. [18]. The von Neumann lattice states are given by

$$|\alpha_{mn}\rangle = \hat{D}(\alpha_{mn})|0\rangle, \tag{5}$$

$$\alpha_{mn} = \frac{1}{\lambda\sqrt{2}}\left(mb + i\frac{2\pi\lambda^2}{b}n\right) \quad (m, n \in \mathbb{Z}). \tag{6}$$

We construct infinitely many nonequivalent von Neumann lattice sets of the states by displacing $|\alpha_{mn}\rangle$ as $\hat{D}(\alpha_{mn})\hat{D}(\beta)|0\rangle$ with

$$\beta = \frac{1}{\lambda\sqrt{2}}\left(\bar{X} + i\frac{\lambda^2}{\hbar}\bar{P}\right), \tag{7}$$

where $\bar{X}$ and $\bar{P}$ are real variables confined in the unit cell, $\delta$, located at the origin of the von Neumann lattice

$$\delta: -\frac{b}{2} \leq \bar{X} \leq \frac{b}{2}, \ -\frac{\pi \hbar}{b} \leq \bar{P} \leq \frac{\pi \hbar}{b}. \tag{8}$$

This cell has the area $2\pi\hbar$. Then, the following relations of crucial importance can be shown [18]:

$$\frac{1}{2\pi\hbar}\iint_\delta d\bar{X}\, d\bar{P}\, \langle 0|\hat{D}^\dagger(\beta)\hat{D}^\dagger(\alpha_{mn})\hat{D}(\alpha_{m'n'})\hat{D}(\beta)|0\rangle = \delta_{mm'}\delta_{nn'}, \tag{9}$$

$$\frac{1}{2\pi\hbar}\sum_{m,n}\iint_\delta d\bar{X}\, d\bar{P}\, \hat{D}(\alpha_{mn})\hat{D}(\beta)|0\rangle\langle 0|\hat{D}^\dagger(\beta)\hat{D}^\dagger(\alpha_{mn}) = 1. \tag{10}$$

The probability distribution function associated with a normalized state, $|\psi\rangle$, on the von Neumann lattice is defined by

$$p_\psi(m,n) = \frac{1}{2\pi\hbar}\iint_\delta d\bar{X}\, d\bar{P}\, \left|\langle 0|\hat{D}^\dagger(\beta)\hat{D}^\dagger(\alpha_{mn})|\psi\rangle\right|^2. \tag{11}$$

From Eq. (10), $p_\psi(m,n)$ is found to be normalized:

$$\sum_{m,n} p_\psi(m,n) = 1. \tag{12}$$

This property should be compared with that of the unaveraged quantity, $Q(m,n) \equiv |\langle \alpha_{mn} | \psi \rangle|^2$, which does not sum up to unity, as is well known. In the case of a mixed state described by the density matrix, $\hat{\rho}$, Eq. (11) is to be modified as follows:

$$p_\rho(m,n) = \frac{1}{2\pi\hbar} \iint_\delta d\bar{X}\, d\bar{P}\, \langle 0 | \hat{D}^\dagger(\beta) \hat{D}^\dagger(\alpha_{mn}) \hat{\rho}\, \hat{D}(\alpha_{mn}) \hat{D}(\beta) | 0 \rangle. \tag{13}$$

Now, taking advantage of the above construction, we present the following new entropy defined on the von Neumann lattice:

$$S[\psi] = -\sum_{m,n} p_\psi(m,n) \ln p_\psi(m,n). \tag{14}$$

This quantity can be thought of as a measure of uncertainty associated with simultaneous measurements of the position and momentum observables on the von Neumann lattice points. We discuss, in what follows, some basic properties of this entropy.

First of all, we wish to point out that the continuous translation symmetry is manifestly broken in $S[\psi]$, in contrast to the Wehrl entropy. That is, $S[\psi]$ is not invariant under the transformation, $|\psi\rangle \to \hat{D}(z)|\psi\rangle$ ($z \in \mathbb{C}$). However, it is still invariant under the discrete translation on the lattice subspace, i.e., the translation that can be absorbed by trivial shift of the summation over all integers.

To our present knowledge, it is not possible to analytically derive the rigorous value of the lower bound of $S[\psi]$. Therefore, it seems legitimate to evaluate it for a class of the coherent states, $|\psi\rangle = |z\rangle = \hat{D}(z)|0\rangle$ ($z \in \mathbb{C}$). The corresponding probability distribution function on the von Neumann lattice, $p_z(m, n)$, is calculated to be

$$p_z(m, n) = \frac{1}{2\pi \hbar} \iint_\delta d\bar{X}\, d\bar{P}\, \exp\left[-\left|\beta + (\alpha_{mn} - z)\right|^2\right]$$

$$= \frac{1}{4} \mathrm{Erf}\left(\rho_m - \frac{b}{2\sqrt{2}\,\lambda}, \rho_m + \frac{b}{2\sqrt{2}\,\lambda}\right)$$

$$\times \mathrm{Erf}\left(\sigma_n - \frac{\pi\lambda}{\sqrt{2}\,b}, \sigma_n + \frac{\pi\lambda}{\sqrt{2}\,b}\right). \tag{15}$$

where

$$\rho_m = \frac{mb}{\sqrt{2}\,\lambda} - \mathrm{Re}\,z, \qquad \sigma_n = \frac{n\pi\sqrt{2}\,\lambda}{b} - \mathrm{Im}\,z, \tag{16}$$

and $\mathrm{Erf}(x_1, x_2)$ is the error function defined by

$$\mathrm{Erf}(x_1, x_2) = \frac{2}{\sqrt{\pi}} \int_{x_1}^{x_2} dx\, e^{-x^2}. \tag{17}$$

The associated entropy is given by

$$\mathsf{S}[z] = -\sum_{m,n} p_z(m, n) \ln p_z(m, n). \tag{18}$$

It should be noted that, from the intergal representation in Eq. (15), it follows that

$$p_z(-m, -n) = p_{-z}(m, n), \tag{19}$$

which implies

$$\mathsf{S}[z] = \mathsf{S}[-z]. \tag{20}$$

We have already mentioned that $\mathsf{S}[z]$ explicitly depends on the coherent state parameter, $z$, due to the broken continuous translation invariance. Here, we seek the stationarity condition on $\mathsf{S}[z]$ with a given fixed value of

$$c \equiv \frac{b}{\sqrt{2\pi}\,\lambda}. \tag{21}$$

Though the detailed calculations performed are not present here, both $\partial S[z]/\partial \mathrm{Re}\,z$ and $\partial S[z]/\partial \mathrm{Im}\,z$ can be shown to vanish when $z=0$. Thus, $S[z=0]$ is a candidate of the extremum of $S[z]$. Numerical evaluation tells us that $S[z=0]$ is actually the stationary value of $S[z]$. This suggests that $S[\psi]$ takes its minimum for the ground state of the harmonic oscillator.

The quantity, $S[z=0]$, still depends on the parameter, $c$, given in Eq. (21). Let us examine for what value of $c$ it becomes the global minimum. Using the integral representation in Eq. (15) with $z=0$, we find that the derivative of $S[z=0]$ with respect to $c$ vanishes when $c=1$, that is,

$$b = \sqrt{2\pi}\,\lambda. \tag{22}$$

Again, numerical evaluation shows that $S[z=0]$ in fact takes its minimum value when Eq. (22) holds. It is calculated to be

$$S_{\min}[z=0] = 1.386.... \tag{23}$$

Thus, we present the following conjecture:

$$S[\psi] \geq 1.386.... \tag{24}$$

Note that this lower bound is smaller than that of the Wehrl entropy, $1 + \ln \pi = 2.144....$

Finally, we wish to mention that the minimum entropy condition in Eq. (22) is of interest since *it makes the von Neumann lattice perfectly square*, as can be seen from Eq. (6). This means that the present entropy is made increased by any squeezing operations on the lattice. In this sense, it can also be regarded as a measure of classical nature of a quantum state, like the Wehrl entropy.

In conclusion, we have presented a new type of entropy defined on the von Neumann lattice and have discussed its basic properties. We have given a physical interpretation to this quantity as a measure of uncertainty associated with discrete simultaneous measurements of the position and momentum observables on the von Neumann lattice. We have evaluated its values for a class of the coherent states, and have found that the entropy takes its minimum value for the ground state of the harmonic oscillator when the lattice is perfectly square.

**Acknowledgments**

One of us (S. A.) would like to thank Professors A. Mann and M. Revzen for discussions. He also acknowledges the supports from the Grant-in-Aid for Scientific Research of Japan Society for the Promotion of Science and from Institute of Theoretical Physics at Technion-Israel Institute of Technology.